\documentclass[fleqn,10pt]{wlscirep}
\usepackage{algorithm,algpseudocode}
\title{Combining Support Vector Machine and Elephant Herding Optimization for Cardiac Arrhythmias}

\author[1,*]{Aboul Ella Hassanien}
\author[2,*]{Moataz Kilany}
\author[2]{Essam H. Houssein}

\affil[2]{Faculty of Computers and Information, Information Techology Department, Cairo University, Egypt}
\affil[3]{Faculty of Computers and Information, Computer Science Department, Minia University, Egypt}
\affil[4]{Scientific Research Group in Egypt (SRGE) http://www.egyptscience.net}
\affil[*]{aboitcairo@gmail.com}

\keywords{Electrocardiograms(ECG), Cardiovascular Diseases, Elephant Herding Optimization (EHO), Support Vector Machines (SVMs), Feature Selection.}

\begin{abstract}
Many people are currently suffering from heart diseases that can lead to untimely death. The most common heart abnormality is arrhythmia, which is simply irregular beating of the heart. A prediction system for the early intervention and prevention of heart diseases, including cardiovascular diseases (CDVs) and arrhythmia, is important. This paper introduces the classification of electrocardiogram (ECG) heartbeats into normal or abnormal. The approach is based on the combination of swarm optimization algorithms with a modified Pan–Tompkins algorithm (MPTA) and support vector machines (SVMs). The MPTA was implemented to remove ECG noise, followed by the application of the extended features extraction algorithm (EFEA) for ECG feature extraction. Then, elephant herding optimization (EHO) was used to find a subset of ECG features from a larger feature pool that provided better classification performance than that achieved using the whole set. Finally, SVMs were used for classification. The results show that the EHO-SVM approach achieved good classification results in terms of five statistical indices: accuracy, 93.31\%; sensitivity, 45.49\%; precision, 46.45\%; F-measure, 45.48\%; and specificity, 45.48\%. Furthermore, the results demonstrate a clear improvement in accuracy compared to that of other methods when applied to the MIT-BIH arrhythmia database. 

\end{abstract}
\begin{document}

\flushbottom
\maketitle
%
%
\thispagestyle{empty}

\section*{Introduction}
\label{Sec:Introduction}
The World Health Organization (WHO) refers to CVDs as the main cause of death around the world. An estimated 17.5 million people died from CVDs in 2012, representing 31\% of all global deaths \cite{WorldHealth2016}. Accordingly, cardiac health research has received substantial attention from researchers, especially those targeting preventive, medical and technological advances. The main interest of researchers in this field is the improvement of traditional cardiovascular-diagnosis technologies.

ECG is a common and vital diagnosis tool for many cardiac disorders and breathing disorders, such as obstructive sleep apnea syndrome, and for monitoring other functional or structural cardiac abnormalities \cite{BortolanWillems1992}. The availability, reasonable cost, simplicity and low risk of ECG have made it a popular technique that has been applied in many research fields during the past two decades. ECG is a non-invasive tool that measures the electrophysiological activity and of the heart and the cardiovascular system \cite{HasanMamun2012} and analysis of heart function. A heartbeat signals has three main characteristic features: the P wave, QRS complex, and T wave. Each feature appears as a distinguishable peak that is repeated in each beat signal. Cardiac arrhythmia detection requires analysis of the morphology, amplitude, and duration of the P, QRS and T peaks. The automation of ECG signal analysis based on the main characteristic P, QRS and T waves is an important research field for several reasons. Physicians depend on these signals to diagnose many cardiac diseases, such as autonomic malfunction, and other vascular, respiratory or even psychological dysfunctions. The automation process involves numerous fields. This paper employs ECG analysis techniques to produce a model to efficiently and accurately detect heartbeats belonging to a set of categories known by cardiologists. To obtain good results, we combined novel optimization techniques with classifier methods to perform heartbeat classification \cite{houssein2017}.

Several recent studies on ECG classification and modeling have been presented. For example, in \cite{shadmand2016new}, temporal features and the hermit function coefficient are extracted from ECG signals as an input vector of the block-based neural network. In \cite{KoraKalva2015}, the bacterial foraging optimization (BFO) and particle swarm optimization (PSO) are combined with neural networks (NNs) for the detection of left and right bundle branch block ECG patterns. In \cite{MoeinLogeswaran2007}, the cutoff frequency of ECG was investigated, and the spectrum of the ECG signal was extracted from four classes. In \cite{PasolliMelgani2015}, the proposed algorithm required approximately 15 min to filter a training set composed of 250 labeled samples. A five-level ECG signal quality classification algorithm using SVM was outlined in \cite{LiRajagopalanClifford2014}.

In \cite{YochumRenaudJacquir2016}, computers in Cardiology applied continuous wavelet transform and Daubechies wavelet to the benchmark MIT-BIH Arrhythmia database. In addition, hybrid firefly and PSO (FFPSO) were combined with NNs to detect bundle branch block in \cite{KoraKrishna2016}. Additionally, PSO with a random asynchronous approach was introduced in \cite{AdamShapiaiMohdTumariEtAl2014}. Finally, in \cite{VafaieAtaeiKoofigar2014}, a fuzzy classifier with a genetic algorithm (GA) was proposed to classify ECG signals more precisely based on a dynamic model of the ECG signal.

The aim of this paper is to present an automatic classification approach for cardiac arrhythmias. The results introduced in this paper \cite{jovic2017classification} show that ECG classification of arrhythmias can be highly accurate. Therefore, these past results serve as motivation to focus on the classification of ECG heartbeat signals into normal or abnormal. Feature extraction and selection techniques play a major role in the domain of signal processing. Therefore, the performance of identification systems depends strongly on these techniques \cite{mihandoost2017cyclic}. This paper introduces a hybrid optimization and classification approach that uses EHO \cite{WangDebCoelho2015} to select relevant features and optimize the SVM classifier parameters for ECG heartbeat signals.

The introduced classification approach is superior to alternative approaches in a number of aspects such as; 1) We applied a recent optimization algorithm that employs a simple and relatively quick search pattern. 2) Our validation relied on a stable benchmark dataset acquired by the MIT-BIH Laboratory and employed a relatively large number of records (10 patients). 3) Validation was conducted using 3-fold leave-one-out cross-validation for generalized performance. 4) The classification model relied on a large number of features compared with previous ECG classification problems. 5) Staged optimization was employed for ECG classification optimization, i.e., feature selection and parameter optimization were performed in separate stages, in contrast to many former studies. Staged optimization prevents the loss of opportunities that arise when search agents change correlated parameters (features and classification penalty) simultaneously.

The structure of this paper is as follows. The next section introduces the techniques and materials employed in this paper. The methodology is explained in detail in terms of the applied dataset, feature extraction and selection, and emotion regression optimization process. Then, the experimental results and a discussion of these results are presented. Finally, concluding remarks and future work are provided in last section.

\section*{Materials and Methods} 
\label{Sec:Methods}

\subsection*{Electrocardiogram (ECG) Signals}\label{Sec:ECG}
Six known heartbeat types can be identified. Each heartbeat can be accurately described by an ECG waveform consisting of five peaks (features). The detection and evaluation of each peak and its variance, distance and other mathematical characteristics leads to a powerful identification of heartbeat properties \cite{KutluKuntalp2011}. Table \ref{TBL:ECGWaves} shows a description of each waveform. All these characteristic points should be detected.

\begin{table}[!ht]
	\centering
	\caption{ECG waves.}
	\label{TBL:ECGWaves}
	\begin{tabular}{|l|l|}
		\hline
		\textbf{Wave} & \textbf{Description}                                              \\\hline 
		P        & A trial depolarization.  \\
		Q        & Point before R, with slope $< 0$.\\
		R        & Distance between two peaks of QRS. \\
		S        & Point after R with slope $> 0$.\\
		T        & Ventricular re-polarization. \\
		\bottomrule
	\end{tabular}
\end{table}

As a part of the ECG automatic detection process, additional features are extracted from the P, Q, R, S and T waveforms as feature vectors \cite{KutluKuntalp2011}. Those five basic components (P, Q, R, S, and T) are used to interpret the ECG, as shown in Table \ref{TBL:ECG_WVFRM_Param}.

\begin{table}[!ht]
	\centering
	\caption{ECG parameters description.}
	\label{TBL:ECG_WVFRM_Param}
	\begin{tabular}{|l|l|}
		\hline
		\textbf{Amplitude} & \textbf{Duration} \\\hline 
		P Wave - 0.25 mV        & PR interval- 0.12s to 0.20s \\
		Q Wave -25\% of R wave & ST interval- 0.05s to 0.15s \\
		R Wave -1.60 mV         & QT interval- 0.35s to 0.44s \\	
		T Wave -0.1 to 0.5 mV   & QRS interval-0.09s          \\\hline 
	\end{tabular}
\end{table}

\subsection*{Elephant Herding Optimization}

A new algorithm introduced by Gai-Ge Wang et. al. in 2012, named Elephant Herding Optimization Algorithm (EHO) \cite{WangDebCoelho2015}. EHO solve all kinds of global optimization problems and the herding behavior of the elephants can be modeled as follow; (1) each population is composed of some clans in the same time each clan has fixed number of elephants. (2) at each generation, a fixed number of male will leave their family group and live far away. (3) in each clan, the elephants live together under the leader called a matriarch. Exploration and exploitation in EHO are achieved by the clan updating operator and the separating operator. Algorithm \ref{alg:EHO} provides the algorithmic framework of the EHO. For more details about EHO, see \cite{WangDebCoelho2015}.

\begin{algorithm}[htb!]
	\caption{Pseudo code of EHO.}
	\label{alg:EHO}
	\begin{algorithmic}[1]
		\State{\textbf{Initialization:} Initialize the generation counter $g=1$; the maximum generation $MaxGen$ and the population; }
		\State{\textbf{While $g<MaxGen$ do}}
		\State All the elephants should be classified according to the fitness (objective function)
		\State Perform clan updating operator
		\State Perform separating operator
		\State assess the population by newly updated positions
		\State $g=g+1$
		\State{\textbf{end while}}
	\end{algorithmic}
\end{algorithm}

\subsection*{Support Vector Machines (SVMs)}
Several classifiers have been proposed in the signal processing domain, including artificial neural network \cite{de2009patient}, SVM, and fuzzy logic system \cite{TsipourasVoglisFotiadis2007}. Most researchers have focused on SVM for CVD classification of ECG signals \cite{GhoshMidyaKoleyEtAl2005}. The classification process of ECG signals for CVDs using SVM is regarded as the main objective of this paper. Previous research illustrated the great performance of SVM, in which data are represented as a P-dimensional vector \cite{KarpagachelviArthanariSivakumar2012}. Classification is performed by means of optimal separating hyperplanes, which ensure the greatest margin between the closest data points that belong to separate classes. SVMs depend on kernels in the classification process, and kernel selection is a challenging task that strongly affects the classification performance. The SVM algorithm aims to find the greatest distance around a hyperplane to separate a positive class from a negative class, as illustrated in the following Equations. 

\begin{equation}
\label{EQ:SVM1}
f(x) = (w.\phi(x)) + b
\end{equation}

\begin{equation}
\label{EQ:SVM2}
R_{SVM}(C) = c \frac{1}{N} \sum_{i = 1}^{N} L_\epsilon (y_i y_i^\Delta) + \frac{1}{2} W^T.W
\end{equation}

\begin{equation}
{{L}_{\varepsilon }}({{y}_{i}}y_{i}^{\Delta })=\left\{ \begin{matrix}
\left| {{y}_{i}}-y_{i}^{\Delta } \right|-\varepsilon \left| {{y}_{i}}-y_{i}^{\Delta } \right|\,\,\,,\ge \varepsilon   \\
0,\,\,\,\,\,\,\,\,\,\,\,\,\,\,\,\,\,\,\,\,\,\,\,\,\,\,\,\,\,Otherwise  \\
\end{matrix} \right.
\end{equation}

\begin{equation}
\label{EQ:SVM4}
y^\Delta = f(x) = \sum_{i = 1}^{N}(\alpha_i - \alpha_i^*) K(X_i,X) + b
\end{equation}

\begin{equation}
K(X_i,X) = exp((\frac{-1}{\delta^2} (X_i - X_j)^2))
\label{EQ:SVM5}
\end{equation}
Where $\phi(x)$ is a non-linear high-dimensional feature space and $x$ is the input space. $w$ and $b$ are the modifiable model and threshold, estimated by minimizing, respectively. $\alpha_i - \alpha_i^*$ is a Lagrange multipliers. $k(x_i,x)$ and $\delta^2$ defines Gaussian kernel and the width of the kernel function, respectively. $C$ is a positive real constant. $\epsilon$ refers to SVM parameter.

\subsection*{Proposed Approach for ECG Heartbeat Classification}  
\label{Sec:ProposedApproach} 
\begin{figure*}[!ht]
	\centering
	\includegraphics[width = 0.8\textwidth,height=10cm]{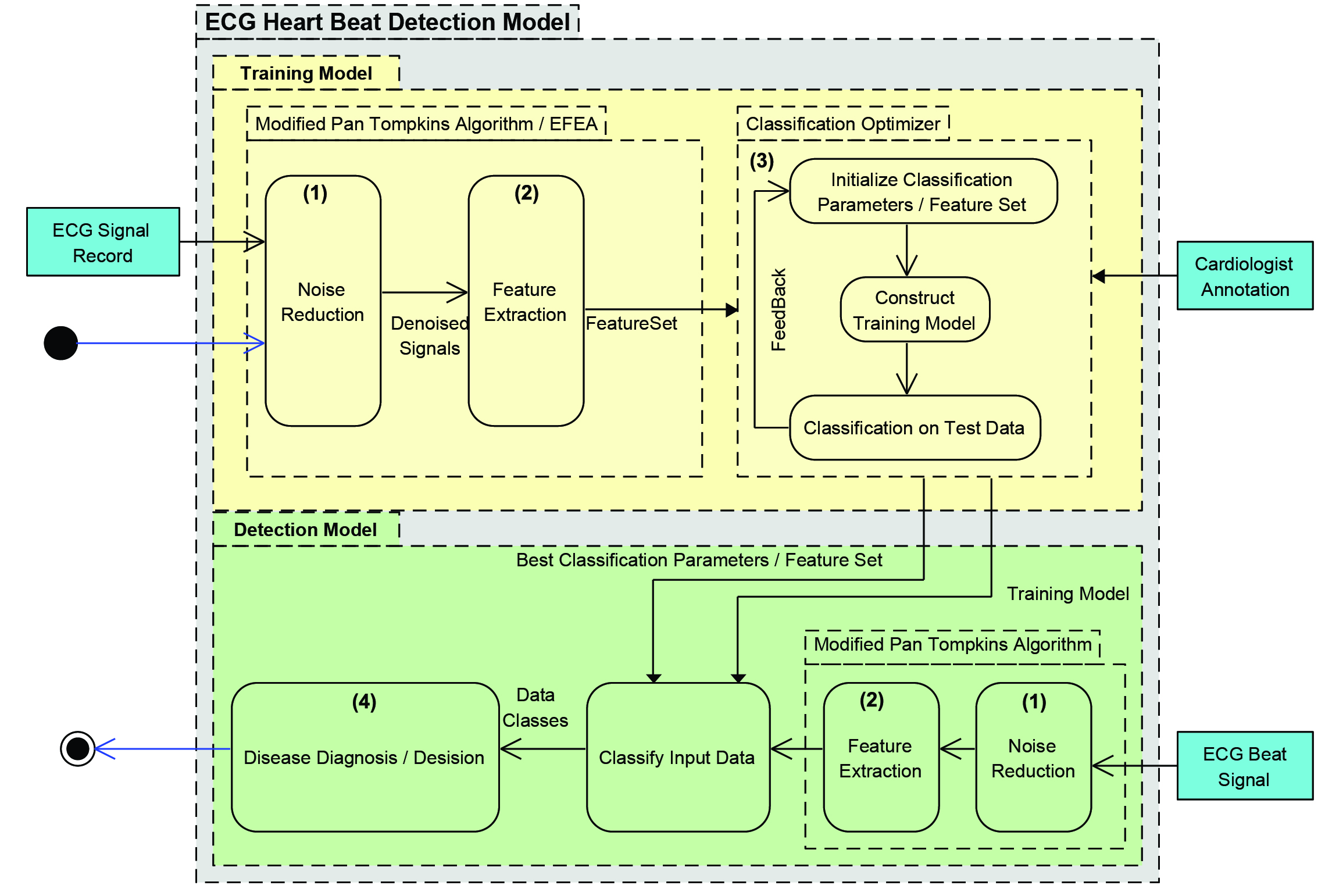}
	\caption{The proposed approach for ECG heartbeat classification.}
	\label{fig:ECGApproach}       
\end{figure*}

\begin{figure*}[!htb]
	\centering
	\includegraphics[width = 0.7\textwidth, height = 11.5cm]{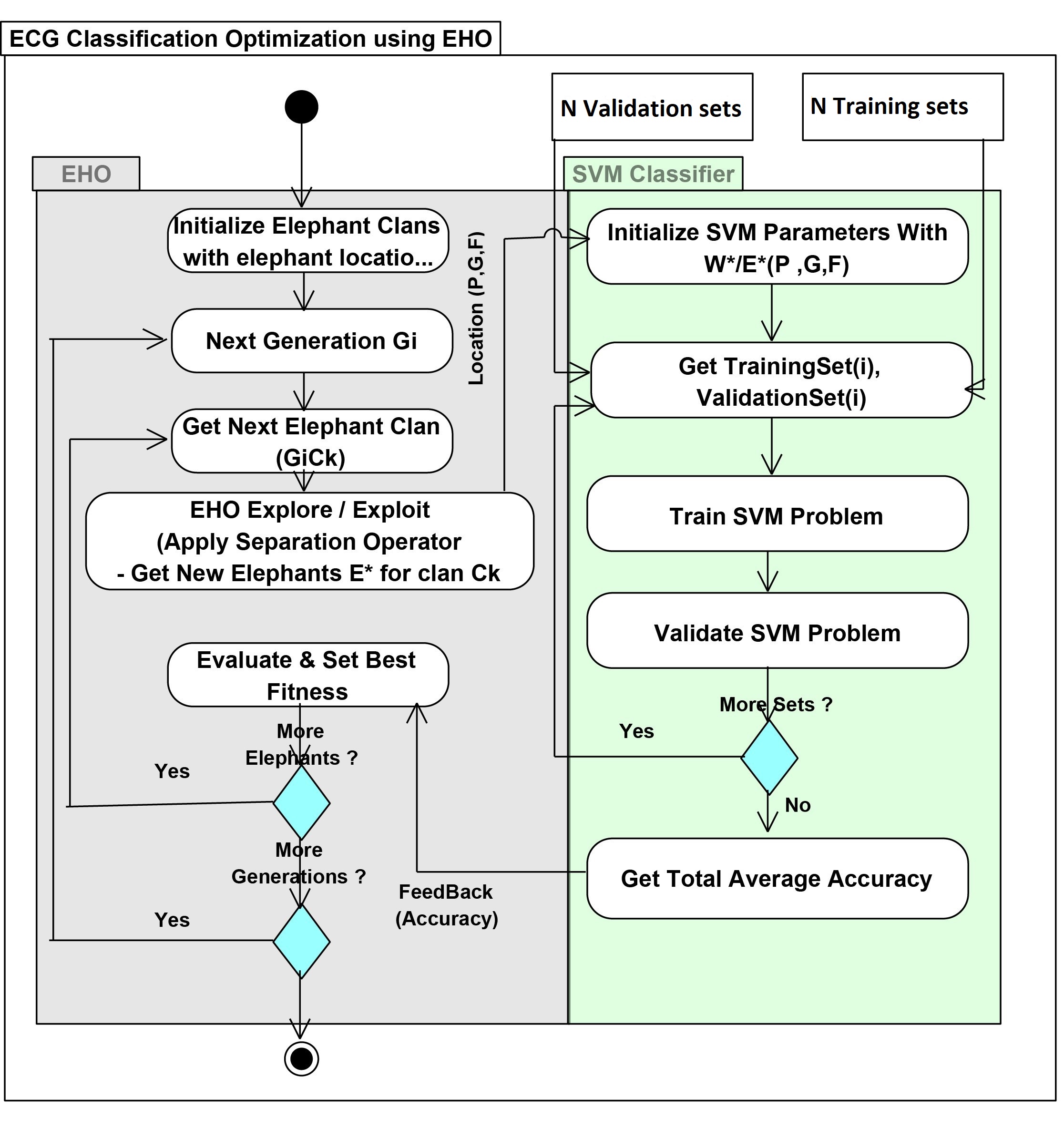}
	\caption{The general approach for ECG heartbeat classification based on EHO.}
	\label{fig:ECG_Optimization}       
\end{figure*}

Five distinct points (P, Q, R, S and T waves) are included in each ECG signal. Fig. \ref{fig:ECGApproach} shows the four phases of the SVM feature optimization process of the proposed approach: (1) Preprocessing, (2) ECG feature extraction, (3) Feature selection and optimization, and (4) Classification and validation. Later, we provide a detailed model for phases (3) and (4), which are shown in Fig. \ref{fig:ECG_Optimization}.

In this paper, the EHO algorithm was modified for the purpose of classification optimization. Elephant locations are identified as SVM parameters in the selected features set while elephant fitness is realized as the average classification accuracy for all cross-validation folds. For the fitness calculation, the SVM is trained with three training sets and validated against three validation sets. Algorithm \ref{alg:EHO-SVM} provides the algorithmic framework of the EHO-SVM classifier presented in Fig. \ref{fig:ECG_Optimization}. Algorithm \ref{Alg:Fitness} shows the fitness calculation process using the SVM classifier.

\begin{algorithm}[htb!]
	\caption{EHO-SVM approach.}
	\label{alg:EHO-SVM}
	\begin{algorithmic}[1]
		\State{Input: Training sets (Folds) ($T1,T2 \dots Tn$)}
		\State{Input: Validation sets (Folds) ($V1,V2 \dots Vn$)}
		\State{Output: Classification accuracy}		
		\State{\textbf {Initialization:}\\
			$Generation~counter~t \gets 1$ \\
			{Initialize population locations (SVM, Kernel parameters / Selected Features)}} \\
			Evaluate population fitness ($g$) (Alg. \ref{Alg:Fitness}, Eq. \ref{EQ:Fitness})
		\State{\textbf{While $g<MaxGen$ do}}
		\State Sort all the elephants according to their fitness
		\State Apply clan updating operator
		\State Apply elephant separating operator
		\State Evaluate population fitness ($g$) (Alg. \ref{Alg:Fitness}, Eq. \ref{EQ:Fitness})
		\State Find best elephant with highest fitness (Classification accuracy)
		\State $g=g+1$
		\State{\textbf{end while}}
	\end{algorithmic}
\end{algorithm}

\subsubsection*{Fitness Function}
An optimization algorithm generally depends on a fitness function to find best solution. The fitness function provides the algorithm a value that quantifies the fitness of each solution found in search space. In this paper, we selected classification accuracy as the solution qualifier through the search process. Classification accuracy is in the range $[0,1]$, and each elephant (search agent) is characterized by a number of accuracies that depend on the cross-validation strategy. In this paper, each elephant has three accuracy values, one for each fold in the 3-fold cross-validation strategy. The accuracy values for all folds are averaged to obtain the fitness value for the search algorithm, as shown in Equation \ref{EQ:Fitness}.

\begin{equation}
\label{EQ:Fitness}
f(i,j) = \frac{\sum\limits_{k = 1}^{n} Acc_{i,j,k}}{n}
\end{equation}
where $f(i,j)$ is the fitness value for elephant $i$ in iteration $j$. $n$ represents the number of folds selected for cross-validation. $Acc_{i,j,k}$ is the accuracy of the evaluation for elephant $i$ in iteration $j$ for the data fold $k$.

Algorithm \ref{Alg:Fitness} shows the fitness calculation process using the SVM classifier.

\begin{algorithm}[htb!]
	\caption{Evaluate elephant population fitness.}
	\label{Alg:Fitness}
	\begin{algorithmic}[1]
		\State{Input: Training sets (Folds) ($T1,T2 \dots Tn$)}
		\State{Input: Validation sets (Folds) ($V1,V2 \dots Vn$)}
		\State{Input: Population number ($j$)}
		\State{Output: Total accuracy}
		\For{Each elephant $i$ in population $j$ $E_{i,j}$}
		\State {Get elephant location $Loc_{i,j} \gets Location(E_{i,j})$}
		\For{Each training set $T_i \in {T1, T2, T3}$}
		\State SVM parameters P $\gets GetParameters (Loc_{i,j})$
		\State $Selected~Features~F \gets GetFeatures(Loc_{i,j})$
		\State Train SVM on $T_i$ using $P,F$
		\State $Validation~Accuracy~V \gets$ Validate $V_i$	\State $Acc \gets Acc + V$
		\EndFor
		\EndFor			
		\State $TotalAccuracy \gets Acc / n$
		\State $Fitness \gets TotalAccuracy$		
		\State Exit
	\end{algorithmic}
\end{algorithm}

\subsubsection*{Pre-processing Phase Using MPTA} \label{subsec:MPTA_PreProcessing} 
\label{subsec:ModifiedMPTA}
Power-line interference and baseline wandering are regarded as the most prominent types of noise that strongly affect signals. Patient respiration, with a frequency in the range of 0.15 to 0.3 Hz, is the main source of baseline wandering. Power-line interference is categorized as narrow-band noise centered at approximately 60 Hz and occupies a bandwidth less than 1 kHz. The other sources of noise are wide-band and also affect ECG signals. The hardware used to acquire ECG signals has the ability to suppress power-line interference; however, wide-band noise and baseline wandering cannot be suppressed by hardware alone. Therefore, software algorithms are used to remove baseline wandering and other wide-band noise \cite{RochaParedesCarvalhoEtAl2010}.

In this paper, MPTA \cite{Kriti2014} is used to remove different types of artifacts and noise. First, a bandpass filter, composed of a low-pass filter and a high-pass filter, is used to reduce noise. Then, a derivative filter is used to obtain the slope information. Amplitude squaring is performed, and the signal is passed to a moving window integrator. Finally, a thresholding technique is applied, and the peaks are detected.

\subsubsection*{ECG Feature Extraction} \label{subsubsec:Extendedtion}
A wave analysis technique is required to perform feature extraction. Wave analysis techniques decompose a given wave into its wavelet building blocks. In this paper, two feature extraction techniques are applied to extract features for classification, such as the RR interval.

\textbf{{Feature Extraction Using MPTA: }}
Nine heartbeat waves are extracted from the ST segment and QRS complex based on MPTA, and the ECG signals are decomposed into low-frequency signals. Therefore, the low-frequency band is utilized to detect the P, QRS, and T waves.

\textbf{{Feature Extraction Using Improved Feature Extraction Algorithm (IFEA): }}
We apply an IFEA \cite{houssein2016two} to obtain more features. The algorithm takes the output of MPTA, pinpoints the wave components from the results, and calculates new features. The MPTA output describes different types of locations (points in time) on the ECG waveform in terms of descriptive letters (annotations). There are numerous letters employed for this purpose such as P, N, and T representing the P-type wave, the R-Peak of a normal beat, and the T-Type wave respectively. There are also auxiliary letters such as the opening ad closing brackets representing the beginning and end of a wave type with the wave peak enclosed in between. The following is an example of three consecutive heartbeats from the data set : (P)(N)(t)(P)(V)(t)(P)(N)(t) which are two normal beats with a Ventricular Arrhythmias beat in the middle. The R-Type wave peak takes a series of letters that annotates the type of heart beat in a whole. For example, N is a normal beat as in the ``(P) (N) (t)'' wave. A ``(P) (V) (t)'' wave form indicates a beat with a Ventricular Arrhythmias. The algorithm also resolves some defects in the P-QRS-T extractor when patterns of the waveforms are not consistent, not complete or do not exist for the corresponding beats.

MPTA is employed to extract the nine heartbeat wave characteristic features (1 through 9) shown in Table \ref{TBL:AdditionalEFEA}, which represent detailed features of the previously described P, Q, R, S, and T waveforms. The additional ten features (10 through 19) are extracted by IFEA. All nineteen features extracted with MPTA and IFEA are depicted in Table \ref{TBL:AdditionalEFEA}.

\begin{table}[!htb]
	\centering
	\caption{Heartbeats features extracted by MPTA and EFEA.}
	\label{TBL:AdditionalEFEA}
	\begin{tabular}{|l|l|l|}
		\toprule
		\textbf{N}  & \textbf{Feature}              & \textbf{Meaning}                    \\ \hline 
		1 & PS      & Beginning location of P wave form. \\
		2 & P       & Peak location of P wave form.      \\
		3 & PE      & End location of P wave form.       \\
		4 & Q       & Beginning of QRS complex.          \\
		5 & R       & R peak of QRS complex.             \\
		6 & S       & End of QRS complex.                \\
		7 & TS      & Beginning of  T wave form.         \\
		8 & T       & Peak of T wave form.              \\ 
		9 & TE      & End of T wave form.                \\	\hline
		10 & QRS             & {QRS = S Q.}               \\
		11 & P-R       		 & {P\_RSeg = Q PE.}          \\
		12 & P-R      		 & {P\_RInt = Q PS.}          \\
		13 & S-T       		 & {S\_TSeg = TS S.}          \\
		14 & Q-T      		 & {TE Q.}           \\
		15 & R-R      		 & {RNext R.}                  \\
		16 & P-P             & {PNext P.}                 \\
		17 & R-R / P-P       & {RR-PPSim = ABS(R-R P-P).} \\
		18 & R-R variance    & {Var (R-R).}                 \\
		19 & Heartbeat       & {60/R-R.}                    \\  \hline 
	\end{tabular}
\end{table}

\subsubsection*{Feature Selection and Optimization Based on EHO}
\label{subsec:WWO-EHO-FeatureOptimization}
Swarm intelligence (SI) is a new branch of artificial intelligence employed to mimic the collective behavior of social swarms in nature, such as elephant herding, social spiders, gray wolves, and ant colonies. A swarm is composed of a set of agents that interact among themselves and with the environment without central control. Recent research introduced swarm-based algorithms that can rapidly solve search-based problems at low cost. The types of swarms include nature-inspired and population-based. Classification and feature optimization(feature selection and parameter optimization) are two of many application domains that successfully employ SI. Other domains include machine learning, bioinformatics, medical informatics, dynamical systems and operations research \cite{HassanienEmary2016}. The proposed approach utilizes EHO for feature selection and parameter optimization to improve the classification accuracy.

The feature optimization framework is illustrated in Fig. \ref{fig:ECG_Optimization}, which depicts the last two phases of the classification approach: Phase 3 is feature selection and optimization and Phase 4 is classification and validation. Fig. \ref{fig:ECG_Optimization} also shows how EHO employs the SVM classifier to evaluate the fitness of each search agent in each optimization iteration.

\subsubsection*{ECG Classification and Optimization Parameters}
Research efforts have shown the dependency between feature optimization and SVM parameter optimization. A known approach is to perform optimization via multiple stages of feature optimization followed by SVM parameter optimization rather than simultaneously optimizing features and parameters in the same run.

\begin{figure}[htb!]
	\centering
	\includegraphics[width = 8cm,height=6cm]{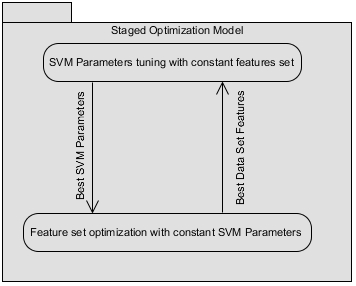}
	\caption{The staged feature and parameter optimization approach.}
	\label{fig:ECG_Stage}       
\end{figure}

Fig. \ref{fig:ECG_Stage} illustrates the multi-stage feature and parameter optimization model. In this work, we established a four-stage optimization process, where the parameter and feature optimization processes are interchanged in each stage.

\section*{Results}
\label{Sec:Results}
The simulation results are obtained using MATLAB R2014a. The experimental setup of the dataset, training data and testing data is presented in the following section.

\subsection*{ECG Dataset Description}
\label{subsec:ECGDataBase}
Researchers use standard databases for analysis purposes. The PhysioNet website is dedicated to medical data corresponding to various diseases \cite{goldberger2000physiobank}. PhysioNet databases are composed of hundreds of digitized medical records of ECG, EEG and other types of physiologic signals. Each ECG record is annotated and revised by a number of cardiologists. Many research efforts depend on the MIT-BIH Arrhythmia database provided by PhysioNet and obtained by the MIT-BIH Laboratory, which consists of several ECG signal records for patients with different types of abnormalities and diseases that affect heart rhythms \cite{KoruerekNizam2010}. MIT-BIH Arrhythmia database comprises of 25 male and 22 female subjects and has 48 half-hours. The signals were collected at 360 samples/sec/channel over a 10 mV range with 11-bit resolution. Additionally, each record is annotated by two or more cardiologists independently, and approximately 110,000 annotations are included in the database \cite{KoruerekNizam2010}.

We applied the proposed classification approach to a subset of the dataset that includes 10 patients with 16 heartbeat types. The data were processed into 10 feature vectors (one for each patient), which combined represent 24,474 records of 10 features each. These data are considered sufficiently large to cover the great variability of patients while maintaining a reasonable level of computational overhead. Table \ref{TBL:ECG_DataSets} shows the datasets employed in our experiment.

\begin{table*}[!ht]
	\centering
	\caption{ECG dataset description.}
	\label{TBL:ECG_DataSets}
	\begin{tabular}{|l|l|l|l|l|}
		\hline
		\textbf{N} &  \textbf{Patient No.} & \textbf{Gender} & \textbf{Age}     & \textbf{PhysioNet Standard Beat Types} \\ \hline 
		1     & 202                   & Male   & 68      & N-A-a-V-F                            \\
		2     & 203                   & Male   & 43      & N-a-V-F                              \\
		3     & 205                   & Male   & 59      & N-A-V-F                              \\
		4     & 207                   & Female & 89      & L-R-A-V-E                            \\
		5     & 214                   & Male   & 53      & L-V-F                                \\
		6     & 215                   & Male   & 81      & N-A-V-F                              \\
		7     & 217                   & Male   & 65      & N-V-/-f                              \\
		8     & 219                   & Male   & Unknown & N-A-V-F                              \\
		9     & 221                   & Male   & 83      & N-V                                  \\
		10    & 223                   & Male   & 73      & N-A-a-V-F-e                          \\   \hline
	\end{tabular}
\end{table*}

The types of heartbeat are represented by symbols defined by PhysioNet, as shown in Table \ref{TBL:BeatsDescription}. This set of beat types is translated from 16 classes into two classes, normal and abnormal (N and A), with type N considered to be normal and all types considered to be abnormal (A). We selected ten patients with a sufficient number of beat types to ensure the validity of classification results and to describe several types of heartbeat. A total of 25,210 ECG beats of different types were used for classification.

\begin{table*}[!ht]
	\centering
	\caption{Heartbeat descriptions.}
	\label{TBL:BeatsDescription}
	\begin{tabular}{|l|l|l|}
		\hline
		\textbf{Beats} & \textbf{Description} & \textbf{Total number} \\ \hline 
		N   & Normal beat     							 & 16742 \\
		L   & Left bundle     							 & 3460  \\
		V   & Premature ventricular contraction   		 & 2154  \\
		/   & Paced beat 								 & 1542  \\
		!   & Ventricular flutter wave					 & 472   \\
		A   & Atrial premature  						 & 228   \\
		f   & Fusion of paced and normal beat			 & 260   \\
		x   & Non-conducted P-wave (blocked APB) 		 & 133   \\
		R   & Right bundle  							 & 86    \\
		$|$ & Isolated QRS-like artifact				 & 37    \\
		F   & Fusion of ventricular                 	 & 30    \\
		a   & Atrial premature 				 & 22    \\
		E   & Ventricular escape 						 & 16    \\
		e   & Atrial escape 							 & 16    \\
		{[} & Start of ventricular flutter/fibrillation  & 6     \\
		{]} & End of ventricular flutter/fibrillation    & 6     \\
		 
		\hline
	\end{tabular}
\end{table*}

\subsection*{Parameters Settings}
\label{subsec:ParametersSetting}
The cross-validation, SVM parameter settings, and EHO parameter settings are included in the experiments. We performed 3-fold leave-one-out cross-validation on all datasets Table \ref{TBL:EHO_SVMParams} summarizes the selected settings for SVM and EHO. A subset of the settings is determined based on the recommendations of the algorithm designers, and the others are set via comprehensive testing.

\begin{table}[htb!]
	\centering
	\caption{Parameter settings for SVMs and EHO.}
	\label{TBL:EHO_SVMParams}
	\begin{tabular}{|l|l|l|l|}
		\hline
		\multicolumn{2}{c}{\textbf{SVM}}             
		&\multicolumn{2}{c}{\textbf{EHO}}  \\ \hline
		\textbf{Parameter}    & \textbf{Value} & \textbf{Parameter}    & \textbf{Value} \\ \hline 
		Kernel                   & Radial Basis  
		&  Alpha Factor          & 5      \\
		Penalty  				& {[}1, 1000{]} 
		& Peta Factor           & 0.0005 \\
		Gamma    				& {[}0, 1000{]} 
		& Elephant Keep 		& 2      \\
		Scaling        			& {[}-1, 1{]}   
		& Clans Count           & 5      \\
		&          		
		& Elephants          	& 30     \\
		\hline
	\end{tabular}
\end{table}

\subsection*{Performance Measurements}
\label{subsec:PerformanceMeasure}

Five standard criteria are used to evaluate the proposed approach: 1) accuracy (Acc), 2) precision (Prec), (3) specificity (Sp), (4) F-measure (F), and (5) sensitivity (Se). Performance measures generally depend on four main metrics of a binary classification result (positive/negative/true/false). Mathematically, the performance measures are defined by the following Equations.

\begin{itemize}
	\item Accuracy (Acc): 
	\begin{equation}
	Acc=\ \frac{TP+TN}{TP+FP+FN+TN}\ *100
	\end{equation}
	
	\item Precision (Prec):
	\begin{equation}
	Prec=\ \frac{TP}{TP+\ FP}\ *  100
	\end{equation}
	
	\item Sensitivity (Se):
	\begin{equation}
	Se=\ \frac{TP}{TP+\ FN}\ * 100
	\end{equation}
	
	\item F-measure (F):
	\begin{equation}
	F=2*\ \frac{PPV*TPR}{PPV+TPR}
	\end{equation}
	
	\item Specificity (Sp):
	\begin{equation}
	Sp=\ \frac{TN}{TN+\ FP}\ * 100
	\end{equation}
\end{itemize}

\section*{Discussion}
\label{Sec:Discussion}
The following section presents the classification results for the 10 selected patients in terms of the performance measures for each patient. Tables \ref{TBL:EHO-Results} and \ref{TBL:EXP_Summary} summarize the classification accuracy, specificity, sensitivity, precision, F-measure for each patient record. The best accuracy results per record are shown in boldface. Each patient has four results sets, one for each stage of the optimization, as discussed in the previous sections. Additionally, the stage in which the best accuracy is obtained for each patient is indicated by boldface font.

\begin{table}[!htb]
	\centering
	\caption{Summary of EHO-SVM approach results.}
	\label{TBL:EHO-Results}
	\begin{tabular}{|l|l|l|l|l|l|l|}
		\hline
		Patient No. & Stage No. & Acc & Prec &Se & F & Sp \\ \hline 
		\multirow{4}{*}{202}     & Stage 1                    
		& 97.09\%     & 44.51\%      & 33.02\%   & 36.19\%      & 85.74\%        \\
		& Stage 2                    
		& 97.10\%     & 19.42\%      & 20.00\%   & 19.71\%      & 80.00\%        \\
		& Stage 3                    
		& 97.10\%     & 19.42\%      & 20.00\%   & 19.71\%      & 80.00\%        \\
		& Stage 4                    
		& \textbf{97.28\%}     & 49.52\%      & 34.03\%   & 38.59\%      & 86.49\%        \\ \hline		
		\multirow{4}{*}{203}        & Stage 1                    
		& 85.95\%     & 35.36\%      & 21.43\%   & 21.13\%      & 81.41\%        \\
		& Stage 2                    
		& 89.68\%     & 17.94\%      & 20.00\%   & 18.91\%      & 80.00\%        \\     
		& Stage 3                    
		& \textbf{89.69\%}    & 24.63\%      & 20.34\%   & 19.61\%      & 80.31\%        \\
		& Stage 4                    
		& 89.69\%     & 24.63\%      & 20.34\%   & 19.61\%      & 80.31\%        \\  \hline		
		\multirow{4}{*}{205}     & Stage 1                    
		& \textbf{98.79\%}     & 47.77\%      & 45.12\%   & 45.92\%      & 92.31\%        \\
		& Stage 2                    
		& 98.79\%     & 47.77\%      & 45.12\%   & 45.92\%      & 92.31\%        \\
		& Stage 3                    
		& 98.72\%     & 47.48\%      & 44.71\%   & 45.42\%      & 91.93\%        \\
		& Stage 4                    
		& 98.76\%     & 47.92\%      & 44.45\%   & 45.80\%      & 91.74\%        \\  \hline
		\multirow{4}{*}{207}     & Stage 1                    
		& 81.21\%     & 32.92\%      & 31.04\%   & 28.41\%      & 83.89\%        \\
		& Stage 2                    
		& \textbf{82.07\%}     & 44.10\%      & 33.65\%   & 32.83\%      & 83.75\%        \\
		& Stage 3                    
		& 78.35\%     & 15.69\%      & 19.97\%   & 17.57\%      & 79.98\%        \\
		& Stage 4                    
		& 80.91\%     & 20.77\%      & 20.32\%   & 18.88\%      & 80.88\%        \\  \hline		
		\multirow{4}{*}{214}     & Stage 1                   
		& 95.66\%     & 46.27\%      & 43.91\%   & 44.50\%      & 93.93\%        \\
		& Stage 2                    
		& 97.21\%     & 48.60\%      & 50.00\%   & 49.29\%      & 50.00\%        \\
		& Stage 3                    
		& 97.21\%     & 48.60\%      & 50.00\%   & 49.29\%      & 50.00\%        \\
		& Stage 4                    
		& \textbf{97.70\%}     & 48.20\%      & 45.98\%   & 46.96\%      & 96.00\%        \\  \hline		
		\multirow{4}{*}{215}     & Stage 1                    
		& 98.75\%     & 47.13\%      & 46.06\%   & 46.57\%      & 95.83\%        \\
		& Stage 2                    
		& \textbf{98.81\%}     & 47.41\%      & 46.08\%   & 46.71\%      & 95.85\%        \\
		& Stage 3                    
		& 98.81\%     & 47.41\%      & 46.08\%   & 46.71\%      & 95.85\%        \\
		& Stage 4                    
		& 98.81\%     & 47.41\%      & 46.08\%   & 46.71\%      & 95.85\%        \\  \hline		
		\multirow{4}{*}{217}     & Stage 1                    
		& 84.95\%     & 71.33\%      & 69.30\%   & 67.63\%      & 94.44\%        \\
		& Stage 2                    
		& 86.06\%     & 74.70\%      & 70.72\%   & 69.19\%      & 94.83\%        \\
		& Stage 3                    
		& \textbf{86.11\%}     & 74.74\%      & 71.13\%   & 69.42\%      & 94.92\%        \\
		& Stage 4                    
		& 86.11\%     & 74.74\%      & 71.13\%   & 69.42\%      & 94.92\%        \\  \hline		
		\multirow{4}{*}{219}     & Stage 1                    
		& 98.14\%     & 44.62\%      & 42.79\%   & 43.47\%      & 90.85\%        \\
		& Stage 2                    
		& 98.70\%     & 49.12\%      & 42.64\%   & 45.30\%      & 90.60\%        \\
		& Stage 3                    
		& \textbf{99.26\%}     & 48.57\%      & 48.36\%   & 48.37\%      & 95.84\%        \\
		& Stage 4                    
		& 99.26\%     & 48.57\%      & 48.36\%   & 48.37\%      & 95.84\%        \\  \hline		
		\multirow{4}{*}{221}     & Stage 1                    
		& 99.54\%     & 99.13\%      & 99.23\%   & 99.18\%      & 99.23\%        \\
		& Stage 2                    
		& 99.67\%     & 99.11\%      & 99.70\%   & 99.41\%      & 99.70\%        \\
		& Stage 3                    
		& \textbf{99.71\%}     & 99.14\%      & 99.83\%   & 99.48\%      & 99.83\%        \\
		& Stage 4                    
		& 99.71\%     & 99.14\%      & 99.83\%   & 99.48\%      & 99.83\%        \\  \hline
		\multirow{4}{*}{223}     & Stage 1                    
		& 88.82\%     & 28.14\%      & 28.11\%   & 27.99\%      & 93.59\%        \\
		& Stage 2                    
		& 90.55\%     & 30.16\%      & 28.68\%   & 29.20\%      & 93.68\%        \\
		& Stage 3                    
		& 90.93\%     & 29.70\%      & 29.59\%   & 29.55\%      & 94.50\%        \\
		& Stage 4                    
		& \textbf{91.81\%}     & 29.26\%      & 30.92\%   & 30.05\%      & 95.91\%  \\     \bottomrule
	\end{tabular}

\end{table}

The problem considered in this paper is not a binary classification problem, so we extract the true positive (TP), false positive (FP), true negative (TN), and false negative (FN) measures by means of a confusion matrix constructed for the classification test.

Table \ref{TBL:EXP_Summary} summarizes the best results for each classifier along with the classification accuracy, precision, sensitivity, F-measure, and specificity. The table compares between accuracy values of SVM and EHO-SVM. The accuracy values of SVM were acquired from early stages of optimization (ST1), where SVM model was assigned random parameters for all patients. Then results were averaged over all patients for each accuracy metric. EHO-SVM accuracy values are calculated as the average of best values for all patients and for each metric stated in Table \ref{TBL:EHO-Results}.

\begin{table}[!htb]
	\centering
	\caption{Summary of the experimental results.}
	\label{TBL:EXP_Summary}
	\begin{tabular}{|l|l|l|l|}
		\hline
		\textbf{Measures} & \textbf{SVM} & \textbf{EHO-SVM} & \textbf{Improvement} \\ \hline 
		Accuracy                   & 80.31\%      & 94.07\%      & 13.76\%              \\
		Precision                  & 40.45\%      & 52.32\%      & 11.87\%              \\
		Sensitivity                & 40.49\%      & 47.85\%      & 7.36\%               \\
		F-measure                  & 38.48\%      & 47.58\%      & 9.10\%               \\
		Specificity                & 40.48\%      & 47.58\%      & 7.10\%               \\ \hline
	\end{tabular}
\end{table}

\begin{figure}[!htb]
	\centering
	\includegraphics[width = 8cm,height=4.5cm]{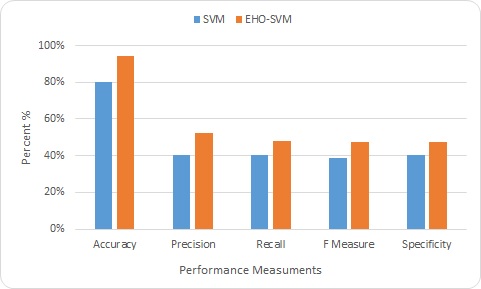}
	\caption{Classification performance for SVM and EHO-SVM.}
	\label{fig:Results}       
	\vspace{0.5cm}
\end{figure}

Fig. \ref{fig:Results} shows the results for each classifier and the visual comparison of the best results obtained by SVM and EHO. The proposed approach achieves the best classification performance with the highest number of features.

MPTA was employed to extract nine heartbeat wave characteristic features (indices 1 through 9), which represent detailed features for the previously described P, Q, R, S, and T waveforms. Additionally, ten features are extracted by means of the proposed IFEA (indices 10 through 19) \cite{houssein2016two}.

\begin{figure}[!htb]
	\vspace{0.3cm}
	\centering
	\includegraphics[width = 0.48\textwidth, height =4.5cm]{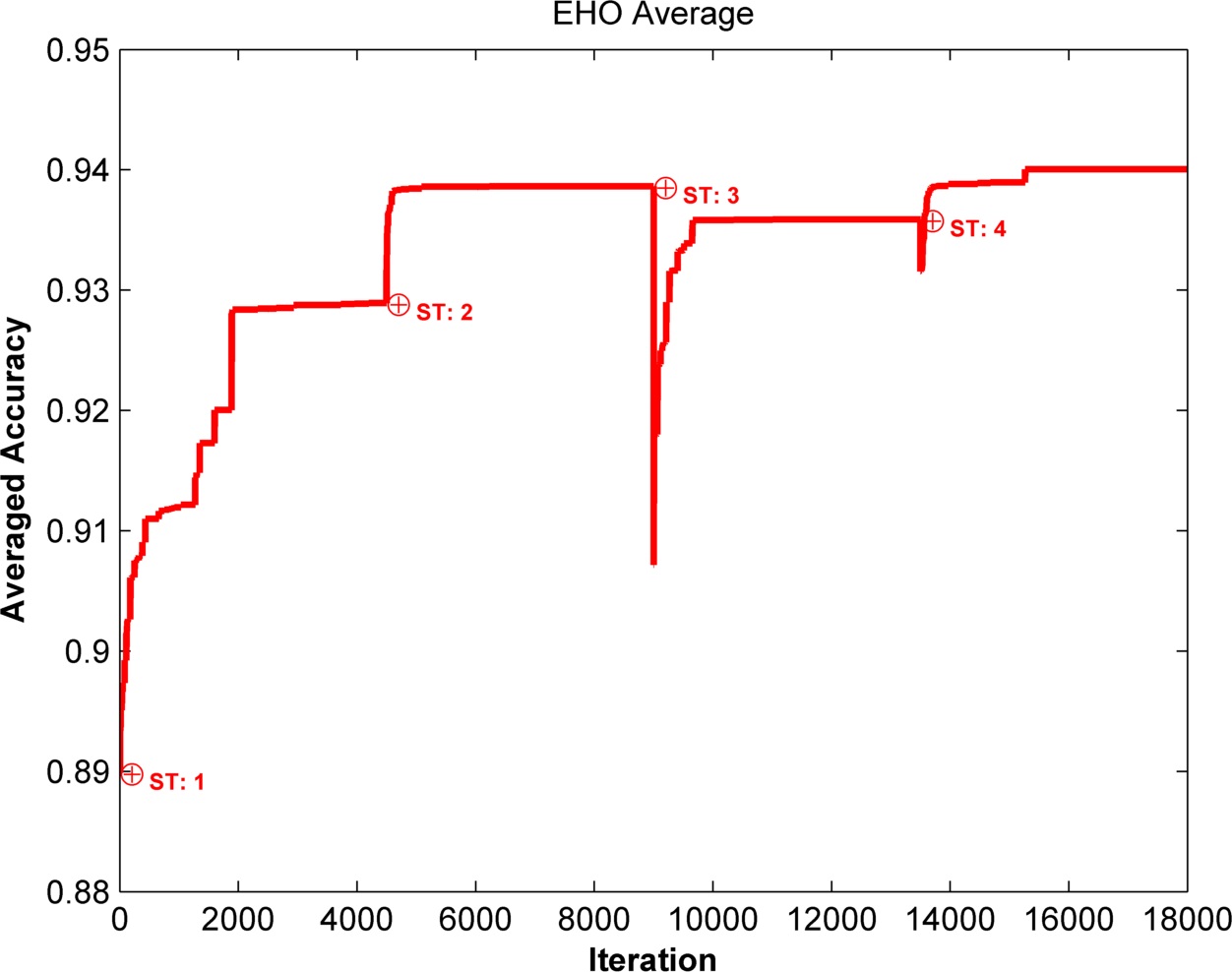}
	\caption{Average convergence curves for EHO.}
	\label{fig:Convergence_EHO}
\end{figure}

The behavior of EHO during the search process is depicted in Fig. \ref{fig:Convergence_EHO}, which illustrates the evolution of the fitness function value (averaged value) associated with the best global swarm parameter for all patients and stages, also known as the convergence of the fitness function, based on EHO.

As shown by Fig. \ref{fig:Convergence_EHO}, each record reaches the maximum classification accuracy at an arbitrary stage. Some records reach the maximum in stage 1, for example, patient number 205 under EHO optimization, while others reach the maximum in stages 2, 3 or 4. Overall, multi-stage optimization is important and can produce better results than those of single-stage optimization.

The convergence curves for EHO show the accuracy of the algorithm and how fast it reaches the final accuracy. For the test conducted in this paper with the defined parameters (Table \ref{TBL:EHO_SVMParams}), EHO reaches the maximum accuracy with fewer iterations. The previous conclusion is valid with respect to both each stage individually and to the overall convergence curve for all stages.

\subsection*{Accuracy Analysis}
To avoid possible bias in the selection of the test and training sets, 3-fold cross-validation is utilized in this paper; hence, the ECG dataset was divided into three parts. For comparison, we consider some previous studies based on the same dataset. In \cite{KoraKalva2015}, the MIT-BIH Arrhythmia database was tested using PSO, GA, BFO, and bacterial foraging–particle swarm optimization (BFPSO) with SVM. In \cite{YochumRenaudJacquir2016}, CWT and the histogram representation were applied to determine the QRS, T and P waves. Furthermore, in \cite{marquez2015study}, the optimal number of Hermite functions to represent the QRS wave was studied. In \cite{chen2017heartbeat}, SVM was utilized to cluster heartbeats based on only two types of features, in contrast to our work with nineteen features. Additionally, in \cite{chen2017heart}, wavelet time frequency (WTF) was applied to detect sudden amplitude and frequency jumps, but the ECG signals were recorded under hypnosis to obtain heart rate variability. However, this work did not focus on the heart rate classification accuracy. These comparisons are shown in Table \ref{TBL:CompareWork}, where it is clear that the proposed approach outperforms the compared studies. The proposed classification approach was applied to 10 patients, 16 types of heartbeat and 24,474 records. The proposed classification approach was validated and evaluated for efficiency based on the sufficiently large data covering a large variety of patients.

It is important to note that this approach requires a number of future improvements. The proposed model currently targets two classes of heartbeat arrhythmias, normal and abnormal, which are considered to be relatively general classes. However, we are improving this work to accurately separate more precise classes of heartbeat, such as PVC, F, A, R, and F. The proposed model also applies a relatively traditional classification technique (SVM); we plan to employ deep learning techniques to achieve better classification performance. Finally, a more advanced and more popular feature extraction technique, such as wavelet transform, is required in future work.

\begin{table}[!htb]
	\centering
	\caption{Comparison of the results and methods of studies that used the same MIT-BIH Arrhythmia database.}
	\label{TBL:CompareWork}
	\begin{tabular}{|l|l|l|l|}
		\hline
		\textbf{Studies} & \textbf{Year} &\textbf{Approach} & \textbf{Accuracy} \\ \hline 
		\cite{marquez2015study} &2015 &Hermite functions   &90\% \\
		\cite{KoraKalva2015} &2015&BFPSO-SVM   &76.74\% \\
		\cite{YochumRenaudJacquir2016} &2016&Delineation Method  &92.44\% \\
		\cite{chen2017heartbeat} &2017&SVM &93.1\% \\
		\cite{chen2017heart} &2017&WTF &NA \\
		Proposed    &2017&EHO-SVM  &93.31\% \\
		\hline
	\end{tabular}
\end{table}

\subsection*{Conclusion and Future Work}
\label{Sec:Conclusion}
ECG analysis helps cardiologist to make decisions about cardiac arrhythmias more accurately and easily to save lives of thousands of people. ECG records the electrical activity of the heart within a specific time; hence, ECG is considered to be an important diagnostic tool to assess heart function. In this paper, we have developed a hybrid approach for automatic ECG signal classification by means of EHO and SVMs. The proposed approach includes three modules for automatic ECG signal classification: an efficient preprocessing module, a feature extraction module, and a feature optimization and classification module. In the preprocessing module, the MPTA and IFEA are applied to extract nineteen heartbeat features. Additionally, we use SVMs to classify features extracted from the previous module. Finally, in the last module, EHO is utilized to optimize the features and parameters extracted by the SVMs. The experiments showed that the proposed approach achieves precise detection. Moreover, the proposed approach shows promise for use by medical experts who wish to diagnose heart and cardiac disorders based on ECG signals. 

In future work, we will propose an automated cardiac arrhythmia classification approach using hybrid SVMs and spike neural network with recent meta-heuristic optimization algorithms to focus on common disorders, such as congestive heart failure, and other cases of biomedical time series. Additional important goals are to analyze ECG signals in time domain and to detect the optimal representation of the P, QRS and T patterns. 


\bibliography{sample}

\section*{Author contributions statement}
 Moataz Kilany performed the experiments, discussed the data and wrote the paper.    Aboul Ella Hassanien conceived and supervised the research, discuss the experiments  and polished the paper.  Essam H. Houssein participated in written some part in the paper and write the references. All authors read and approved the final paper.

\section*{Additional information}
The authors declare no competing interests.

\end{document}